# Challenges in constructing genetic instruments for pharmacologic therapies

Brian A Ference MD, MPhil, MSc; George Davey Smith MD, DSc; Michael V Holmes MBBS, PhD; Alberico L Catapano PhD; Kausik K Ray MD, MPhil; Stephen J. Nicholls MBBS, PhD;

From the Centre for Naturally Randomized Trials, University of Cambridge, Cambridge, U.K. (B.A.F.); MRC Integrative Epidemiology Unit, University of Bristol, Bristol, U.K. (G.D.S.); MRC Population Health Research Unit and the Clinical Trial Service Unit and Epidemiological Studies Unit, Nuffield Department of Population Health, University of Oxford, Oxford, UK (M.V.H.); Department of Pharmacological and Biomolecular Sciences, University of Milan and Multimedica IRCCS, Milano, Italy (A.L.C.); Imperial Centre for Cardiovascular Disease Prevention, Department of Primary Care and Public Health, School of Public Health, Imperial College London, London U.K. (K.K.R.); Monash Cardiovascular Research Centre, Monash University, Melbourne, Australia (S.J.N.)

Correspondence:
Brian A. Ference, M.D., M.Phil., M.Sc., F.A.C.C.
Centre for Naturally Randomized Trials
University of Cambridge
2 Worts' Causeway
Cambridge, UK
CB1 8RN
P: +44 (0)1223 748647
baf29@medschl.cam.ac.uk

**Introduction**

Mendelian randomization studies are increasingly being used to evaluate the potential efficacy and safety of medical therapies as a strategy to provide genetic target validation.[1-7] The first step in this process is to create an instrumental variable consisting of variants in or around the gene that encodes the target of a therapy that are associated with changes in a biomarker that reflects the pharmacologic action of the therapy.

The genes that encode the targets of some therapies include large-effect loss-of-function (or partial loss-of-function) variants that can be used "instrument" the effect of a therapy.[8] If such a variant is associated with a quantitatively large change in the biomarker that reflects the pharmacologic effect of a therapy directed against the target encoded by the gene, and is associated with a corresponding improvement in a clinically relevant outcome measure, then this variant can provide genetic target validation for the therapy under study. Indeed, developing therapies that have genetic target validation doubles the success rate of therapies that reach clinical trials, from 5-10% to 10-20%.[9] While this may fundamentally improve the cost structure of drug development, it still means that 80-90% of therapies directed against genetically validated targets will fail in clinical development!

The reason for this startling expected failure rate of therapies directed against "genetically validated targets" is that genetic target validation is not enough to ensure that a therapy directed against the target will produce a clinically meaningful reduction in the risk of the outcome of interest. Instead, instead genetic target validation is analogous to early stage drug development. This can be understood intuitively by recognizing that large-effect variants tend to be rare. As a result, the estimate of the quantitative magnitude of the observed effect must be viewed with caution because it is generally imprecise with vary wide confidence intervals.

To fill the "translational gap" between discovering genetically validated targets and informing the clinical development of those therapies, it is necessary to conduct a number of additional steps beyond genetic target validation to use naturally randomized genetic evidence to estimate the potential clinical benefit of a therapy directed against a genetically validated target. This can be accomplished by combining two or more common variants in or around the gene encoding a therapy into a genetic score constructed as an instrumental variable. If the common variants included in the instrumental variable genetic score are associated with at least a moderate effect on changes in the biomarker reflecting the pharmacologic action of the therapy it is being used to instrument, then such a genetic score can be used to establish what the causal biomarker is (from a list of biologically related biomarkers altered by the therapy).[10,11] If this type of type of genetic instrument is robust enough, it can also be used to assess the magnitude and shape of the causal dose response curve.[6,10] This information is critical for determining how much the causal biomarker must be altered to produce clinically important effect in a short-term trial to estimate the potential clinical relevance of a therapy directed against the therapeutic target thus guiding clinical development and informing the design of randomized trials evaluating the therapy.

Unfortunately, the genes that encode the target of the vast majority of available therapies do have either rare variants that are associated with a large effect on the biomarker of interest, or any common variants that have at least a moderate effect on that biomarker. In this case, multiple common variants in or around the gene encoding the target of the therapy that have a weak effect on the biomarker reflecting the pharmacologic action of the therapy must be combined into an instrumental variable genetic score. The optimal method to construct an instrumental variable genetic score consisting of several weak effect common variants designed to "instrument" a therapy is unclear. Here, we describe one strategy used to overcome this problem in the context of a recently published study, with updated analyses and results.[7]

*A strategy to overcome challenges in constructing genetic instruments for pharmacologic therapies*

ATP citrate lyase (ACL) is an enzyme located upstream to HMG-CoA reductase in the cholesterol biosynthesis pathway.[12] Inhibiting ACL with bempedoic acid has been shown to reduce plasma LDL-C in several early phase randomized clinical trials.[13-15] Thus, ACL is a biologically validated therapeutic target.

Furthermore, numerous randomized trials have demonstrated that lowering plasma LDL-C by inhibiting the cholesterol biosynthesis pathway by inhibiting HMG-CoA reductase with a statin, NPC1L1 with ezetimibe, or PCSK9 with an antibody reduces plasma LDL-C and reduces the risk of cardiovascular events.[16-20] Thus, the cholesterol biosynthesis pathway is a clinically validated therapeutic target.

The objective of our study therefore was not to provide "genetic target validation" for *ACLY*, the gene that encodes ACL, but instead to construct a genetic instrument that mimics the clinical effects of inhibiting ACL to provide a biological context for interpreting the results of prior ACL inhibitor randomized trials, inform the design of future ACL inhibitor randomized trials, and estimate the clinical effect of lowering plasma LDL-C by inhibiting ACL.[7] We have used this approach to construct genetic instruments that mimic the effect of HMGCR, NPC1L1, PCSK9 and CETP inhibitors in the past to successfully anticipate the results of several landmark randomized clinical trials evaluating these agents (IMPROVE-IT, FOURIER, and REVEAL).[3,4,5,16,17,21]

The challenge in conducting the current study was that, unlike for *HMGCR*, *NPC1L1* and *PCSK9* genes, there are no rare or common variants in the *ACLY* gene region that are strongly associated with changes in plasma LDL-C. To solve this problem, we developed an expanded approach to develop genetic instruments that mimic the effect of therapies directed against existing targets. This process involved several steps.

First, we chose to expand the window around the *ACLY* gene by 500 Kb in either direction to include a greater number of candidate genetic variants with weak effects on LDL-C that could potentially be included in our genetic *ACLY* score designed to instrument ACL inhibition. The justification for this decision was two-fold. First, we noted that genetic variants reported to be associated with *ACLY* expression in the Genotype-Tissue Expression (GTEx) project span a region of 1.7 Mb around the *ACLY* gene (970 Kb upstream and 670 Kb downstream).[22] Thus, circumscribing the window around *ACLY* to a smaller subset of this region maximized the probability that we could potentially include variants in our ACL instrument that were associated with both LDL-C and *ACLY* gene expression, even though they were not located physically close to the gene. Second, we noted that of the 33 genes included in this relatively "gene-dense" region spanning 500 Kb on either side of *ACLY* gene, *ACLY* is the only gene in this region that encoded a protein (ACL) that is both a) known to be involved in lipid metabolism (both cholesterol biosynthesis and the intrinsic fatty acid pathway), and b) results in lower plasma LDL-C when inhibited (with the ACL inhibitor bempedoic acid).

Second, because all variants in this region have at most weak effects on LDL-C, we developed a machine learning algorithm to identify and select variants for inclusion in the *ACLY* score. To maximize computational efficiency, the machine learning algorithms designed to identify variants for inclusion in the *ACLY* score used the public summary data provided by the MAGNETIC consortium.[23] The MAGNETIC summary data were used rather than the larger public Global Lipids Genetics Consortium data because MAGNETIC had imputed data available for a much larger number of genetic variants. The objective of the machine learning algorithm was to identify the group of genetic variants in this gene region that - *in combination* – had the strongest association with LDL-C, which is the main pharmacologic effect of ACL inhibition. A list of the genetic variants that *in* *combination* had the strongest association with LDL-C and that were included in the *ACLY* score ultimately used in our study, and their individual marginal effect sizes for LDL-C as reported in the public MAGNETIC data was provided as Table S3 in the appendix to our manuscript (Table S3: *ACLY* variants included in ACL score and associations with plasma LDL-C);[7] and is reproduced here:

**Table 1:** Individual *ACLY* score variants associations with LDL-C in MAGNETIC GWAS

| variant | Exposure allele | Exposure allele frequency | LDL Effect Size (mg/dl) | LDL SE |
|---|---|---|---|---|
| rs34349578 | G | 0.7793 | -0.7495 | 0.4024 |
| rs55674565 | T | 0.3794 | -0.8229 | 0.3246 |
| rs2883233 | C | 0.1388 | -1.697 | 0.5536 |
| rs117335915 | T | 0.9782 | -3.4469 | 1.2869 |
| rs113201466 | G | 0.9498 | -1.8535 | 0.7931 |
| rs62075782 | G | 0.7774 | -0.9837 | 0.4255 |
| rs145940140 | T | 0.9907 | -5.0301 | 2.2423 |
| rs143382920 | A | 0.9931 | -4.1152 | 2.1669 |
| rs117981684 | G | 0.9789 | -3.1242 | 1.1684 |
|  |  |  |  |  |
| Inverse-variance Meta-analysis (one unit change in score) |  | P=6.4E-10 | -1.173 | 0.1898 |

**NB:** MAGNETIC data used for screening in the machine learning algorithm to identify best genetic score

Importantly, because all variants in the *ACLY* gene region have at most a weak statistical association with LDL-C, using a machine learning algorithm to select variants for inclusion into a score based on their weak statistical associations with LDL-C can potentially "overfit" the data by selecting variants that happen to be weakly associated with plasma LDL-C levels by chance alone. Combining several variants that are weakly associated with LDL-C by chance into a combined score could potentially create a score that is strongly associated with LDL-C due to the compounding effect of multiple chance associations.

To address this over-fitting issue, the machine-learning algorithm selected variants for inclusion in the *ACLY* genetic score using the following protocol. First, a random sample of 50% of the variants within a 500 KB window of the *ACLY* gene was sampled. The variants were then ranked by the strength of their association with LDL-C (z score). The variant most strongly associated with LDL-C was selected for inclusion in the candidate *ACLY* score. All variants correlated with the most strongly associated variant at an r2 > 0.3 were then removed from the remaining set of candidate variants. Next, a random sample of 50% of the remaining candidate variants was sampled. The variant with strongest association with LDL-C was then added to the score, and all variants with an r2 > 0.3 with this variant were then removed from the set of candidate variants. This process repeated iteratively until all variants were either selected, removed due to linkage disequilibrium with a selected variant, or were not associated (p > 0.05) with plasma LDL-C in the MAGNETIC NMR GWS. This overall process was then repeated 1000 times to create a portfolio of candidate *ACLY* instruments.

To further minimize the potential for "over-fitting", we did not rely on the machine learning algorithm alone to construct the *ACLY* genetic score. Instead, the construction of the *ACLY* score followed an iterative three stage process. The first step was the machine learning algorithm designed to identify the groups of variants within the defined gene region that – *in combination* - had the strongest association with LDL-C as described immediately above.

Next, each candidate *ACLY* score generated using this machine learning algorithm was then ranked by the strength of the association of the overall score with plasma LDL-C. For the most strongly associated *ACLY* score, the variants included in this "candidate score" were then evaluated in an independent sample using individual participant data (originally dbGAP studies and later supplemented by the larger INTERVAL study data to improve power)[24,25] for evidence of conditionally independent associations with LDL-C at p< 0.05 as described in the appendix of our manuscript.[7] The conditional associations with LDL-C were estimated in a linear regression model where the dependent variable was plasma LDL-C and the independent variables were age, sex, study sample, and 5 principal components of ancestry among participants who were free of cardiovascular disease at baseline and for whom one or more measurements of plasma lipids and lipoproteins was available. The candidate *ACLY* score was rejected unless all variants included in the score were associated with LDL-C at nominal level of significance (p<0.05) conditional on the effect of all other variants included in the score. This process repeated iteratively until the *ACLY* score that had the strongest overall association with LDL-C, and for which each variant included in the score was conditionally associated with LDL-C, was identified.

As stated in the appendix to our manuscript, the conditional associations for each variant with LDL-C were used as weights in constructing the "weighted" genetic *ACLY* scores.[7] The conditional associations with LDL-C for each variant included in the selected *ACLY* score ultimately used in our study is shown in Table 2 below:

**Table 2:** Individual *ACLY* score variants conditional associations with LDL-C using individual participant data (INTERVAL and dbGAP)

| variant | Exposure allele | Exposure allele frequency | LDL Effect Size (mg/dl) | LDL SE |
|---|---|---|---|---|
| rs34349578 | G | 0.779 | -0.94 | 0.45 |
| rs55674565 | T | 0.379 | -1.25 | 0.44 |
| rs2883233 | C | 0.130 | -0.68 | 0.35 |
| rs117335915 | T | 0.978 | -1.83 | 0.66 |
| rs113201466 | G | 0.950 | -0.99 | 0.44 |
| rs62075782 | A | **0.223** | -0.98 | 0.31 |
| rs145940140 | T | 0.991 | -1.97 | 0.52 |
| rs143382920 | A | 0.993 | -1.19 | 0.43 |
| rs117981684 | G | 0.979 | -0.84 | 0.36 |
| | | | | |
| Inverse-variance Meta-analysis (one unit change in score) | | p = 9.2E-15 | -1.064 | 0.1375 |
| Analyses incorporating partial correlation between variants | | p = 3.4E-16 | -1.2121 | 0.1486 |
| | | | | |

**NB:** *In conditional analyses, the rs62075782 allele associated with lower LDL-C was opposite to that in MAGNETIC. Because, the conditional analyses were the primary weighting exposure, the A allele for rs62075782 was defined as the exposure allele (associated with lower LDL-C in conditional analyses) for all subsequent analyses.*

(We note well here that data in Table S3 of our manuscript, and reproduced as Table 1 above, are the LDL-C effect sizes for each variant as reported in the public MAGNETIC database which can be confirmed by a simple look-up of the public data. Therefore, the LDL-C effect sizes reported in Table S3 of the appendix to our manuscript were not the conditional LDL-C effect sizes for each variant as estimated using individual participant data from INTERVAL and dbGAP and presented in Table 2 above.)

Finally, in the third step the provisionally selected "candidate score" had to divide the population into two approximately equal halves (within the range of 40:60, or 60:40) when dichotomized above or below the median value for the unweighted and weighted genetic scores. This third criterion for selecting the best *ACLY* genetic score was necessary to facilitate the planned 2x2 factorial analysis of the separate and combined effects of genetic instruments that mimic ACL, HMGCR and NPC1L1 inhibitors, respectively. This was a very important consideration because in clinical medicine ACL inhibitors will be used almost exclusively in combination with either a statin or ezetimibe. We note that within this construct, the combined effect of the variants included in the *ACLY* genetic score is the "instrument" that mimics ACL inhibition, not necessarily the effect of any individual variant included in the score.

The *ACLY* instrumental variable that fulfilled each of these three criteria was the one taken forward and used in all subsequent analyses.

### *Associations of ACLY score with plasma LDL-C and apoB in UK Biobank*

Release of biomarker data from the UK Biobank, including data on LDL-C and apoB, *released after our study was published* provides an opportunity to determine whether our *ACLY* genetic score serves as an instrument that mimics the effect of ACL inhibition.[26]

First, we note here that the rs62075782 allele associated with lower LDL-C was the A allele (EAF 0.22) in conditional analyses but was the G allele in the summary MAGNETIC data. Because the conditional effect size with LDL-C was defined as the weight for that variant, all analyses used the A allele (EAF 0.22) as the exposure allele for rs62075782. Subsequent analysis using individual participant data from UK Biobank biomarker data released after publication of our manuscript CONFIRMED that the A allele (EAF 0.26 in UK Biobank) is the allele associated with lower LDL-C.

Second, we note that variant rs34349578 did not pass Quality Control in UK Biobank, and therefore we used variant rs36005199 ($r^2$=0.91) instead for all analyses involving UK Biobank participants.

In the UK Biobank, our *ACLY* score is robustly associated with LDL-C. This is the third independent sample in which this score significantly associates with LDL-C. Here it is important to note that the genetic instrument that we created to mimic ACL inhibition by design was the score itself (i.e. the combined effect of the variants included in the score). The association between the *ACLY* genetic score and LDL-C, and marginal and conditional associations of each variant included in the score, is presented in Table 3:

Table 3: Association of *ACLY* score and individual variants marginal and conditional associations with LDL-C in UK Biobank

| variant | Exposure allele | Exposure allele frequency | Univariate LDL-C ES (mg/dl) | Univariate LDL-C SE | Univariate LDL-C P value | Conditional LDL-C ES (mg/dl) | Conditional LDL-C SE | Conditional LDL-C P value |
|---|---|---|---|---|---|---|---|---|
| rs36005199 | G | 0.779 | -0.4949 | 0.0830 | 2.5E-09 | -0.2153 | 0.1056 | 4.1E-02 |
| rs55674565 | T | 0.379 | -0.2591 | 0.0764 | 7.0E-04 | -0.2614 | 0.1031 | 1.1E-02 |
| rs2883233 | C | 0.130 | -0.1603 | 0.0918 | 8.0E-02 | -0.0777 | 0.1021 | 4.5E-01 |
| rs117335915 | T | 0.978 | -0.3567 | 0.2000 | 7.5E-02 | -0.3413 | 0.2038 | 9.5E-02 |
| rs113201466 | G | 0.950 | -0.4531 | 0.1747 | 9.6E-03 | -0.3723 | 0.1752 | 3.4E-02 |
| rs62075782 | A | 0.223 | -0.4454 | 0.0879 | 4.0E-07 | -0.5163 | 0.1032 | 5.7E-07 |
| rs145940140 | T | 0.991 | 1.0014 | 0.4555 | 2.8E-02 | 1.1499 | 0.4711 | 1.5E-02 |
| rs143382920 | A | 0.993 | 0.1374 | 0.2909 | 6.4E-01 | -0.0095 | 0.3023 | 9.8E-01 |
| rs117981684 | G | 0.979 | -0.1507 | 0.2763 | 5.8E-01 | -0.2771 | 0.2786 | 3.2E-01 |
| | | | | | | | | |
| Inverse-variance Meta-analysis | | | -0.3251 | 0.0391 | **9.6E-17** | -0.2592 | 0.0468 | **3.0E-08** |

NB: rs34349578 did not pass QC in UK Biobank, so rs36005199 was used as the nearest proxy ($r^2$=0.91). Data from 351,267 participants of self-identified White ancestry not on current lipid lowering therapy, adjusted for age gender and the first five principal components

As can be seen in Table 3, the ACLY score is robustly associated with LDL-C in the UK Biobank, using both marginal effect size estimates (p=9.6E-17) and conditional effect estimates (p=3.0E-08).

Furthermore, when evaluated as a calculated score for each person in UK Biobank, the *ACLY* score remains robustly associated with plasma LDL-C both in analyses using the score as a continuous variable (p=3.7E-12) and as a dichotomous variable that acts as an unweighted instrument of randomization that divides participants into two groups with *ACLY* scores above or below the median value for the score (p=1.2E-09)

Table 4: Association of *ACLY* genetic scores with LDL-C in UK Biobank

| ACLY genetic score | LDL-C effect size (mg/dl) | LDL-C SE | P value |
|---|---|---|---|
| Unweighted genetic score as continuous variable | -0.2277 | 0.0328 | **3.7E-12** |
| Unweighted genetic score as dichotomous variable (instrument of randomization) | -0.6485 | 0.1067 | **1.2E-09** |

NB: rs34349578 did not pass QC in UK Biobank, so rs36005199 was used as the nearest proxy ($r^2$=0.91). Data from 351,267 participants of self-identified White ancestry not on current lipid lowering therapy, adjusted for age gender and the first five principal components

Furthermore, when combining data from the MAGNETIC, INTERVAL, dbGAP, and UK Biobank samples – all of which independently replicate the association of the instrument (i.e. the genetic score itself rather than the individual variants) the *ACLY* score was strongly associated with LDL-C with an overall mean effect size per unit change in the score of -0.410 mg/dL (SE 0.0369), and a robust p value of 4.1E-29.

In addition, the *ACLY* score was also robustly associated with apoB in the UK Biobank. In all analyses the genetic instrument was robustly associated with lower plasma apoB, with p values ranging from 5.5E-09 to 4.7E-18, as shown in Tables 5 and 6 below:

**Table 5:** Individual *ACLY* score variants marginal and conditional associations with apoB in UK Biobank

| variant | Exposure allele | Exposure allele frequency | Univariate apoB ES (mg/dl) | Univariate apoB SE | Univariate apoB P value | Conditional apoB ES (mg/dl) | Conditional apoB SE | Conditional apoB P value |
|---|---|---|---|---|---|---|---|---|
| rs36005199 | G | 0.779 | -0.3591 | 0.0610 | 4.1E-09 | -0.1453 | 0.0776 | 6.1E-02 |
| rs55674565 | T | 0.379 | -0.2083 | 0.0562 | 2.1E-04 | -0.2122 | 0.0758 | 5.1E-03 |
| rs2883233 | C | 0.130 | -0.1222 | 0.0675 | 7.0E-02 | -0.0577 | 0.0751 | 4.4E-01 |
| rs117335915 | T | 0.978 | -0.1845 | 0.1471 | 2.1E-01 | -0.1549 | 0.1499 | 3.0E-01 |
| rs113201466 | G | 0.950 | -0.3789 | 0.1285 | 3.2E-03 | -0.3209 | 0.1289 | 1.3E-02 |
| rs62075782 | A | 0.223 | -0.3261 | 0.0647 | 4.7E-07 | -0.3898 | 0.0759 | 2.7E-07 |
| rs145940140 | T | 0.991 | 0.4477 | 0.3353 | 1.8E-01 | 0.6209 | 0.3468 | 7.3E-02 |
| rs143382920 | A | 0.993 | -0.0702 | 0.2141 | 7.4E-01 | -0.1728 | 0.2224 | 4.4E-01 |
| rs117981684 | G | 0.979 | -0.1762 | 0.2031 | 3.8E-01 | -0.2658 | 0.2048 | 1.9E-01 |
| **Inverse-variance Meta-analysis** | | | -0.2493 | 0.0288 | **4.7E-18** | -0.2012 | 0.0343 | **5.5E-09** |

**NB:** rs34349578 did not pass QC in UK Biobank, so rs36005199 was used as the nearest proxy ($r^2$=0.91). Data from 351,267 participants of self-identified White ancestry not on current lipid lowering therapy, adjusted for age gender and the first five principal components

**Table 6:** Association of *ACLY* genetic scores with apoB in UK Biobank

| ACLY genetic score | apoB effect size (mg/dl) | apoB SE | P value |
|---|---|---|---|
| Unweighted genetic score as continuous variable | -0.1745 | 0.0241 | **4.5E-13** |
| Unweighted genetic score as dichotomous variable (instrument of randomization) | -0.4775 | 0.0785 | **1.2E-09** |

**NB:** rs34349578 did not pass QC in UK Biobank, so rs36005199 was used as the nearest proxy ($r^2$=0.91). Data from 351,267 participants of self-identified White ancestry not on current lipid lowering therapy, adjusted for age gender and the first five principal components

In addition, the *ACLY* instrument was associated with a ratio of lower aopB to lower LDL-C (apoB:LDL-C) of approximately 0.75, which is consistent with lowering LDL-C by removing LDL particles through up-regulation of the LDL receptor. Therefore, by demonstrating that the *ACLY* score is robustly associated with both LDL-C and concordantly lower apoB, the UK Biobank data demonstrate that the *ACLY* score is instrumenting a gene that lowers LDL-C by lowering apoB particles through the LDL receptor. The only gene in this region known to be associated with lowering LDL-C through the LDL receptor is *ACLY*. Indeed, these data provide further validation for the mechanism of action by which ACL inhibition reduces LDL-C – i.e. by reducing cholesterol biosynthesis leading to an up-regulation of the LDL receptor with subsequent clearing of circulating LDL particles resulting in a reduction in plasma LDL-C levels.

*Associations of ACLY score with cardiovascular events in the UK Biobank*

Release of the biomarker data from UK Biobank was also accompanied with release of updated data on clinical events, including major cardiovascular events and diabetes.

In our manuscript, the associations for cardiovascular outcomes were reported for the approximately 100,000 cases of major coronary and cardiovascular outcomes in the combined data from CARDIoGRAMplusC4D and UK Biobank overall, and the combined data from UK Biobank and dbGAP for the factorial analyses.[7]

As shown in Tables 7 and 8, both the weighted and unweighted *ACLY* score is robustly associated with CV events in the public summary data from 60,000 cases of CAD in the CARDIoGRAMplusC4D consortium studies and the individual participant data from the UK Biobank (using data prior to the recent update):

Table 7: Individual *ACLY* score variants *unweighted* associations with CAD in CARDIoGRAMplusC4D and UK Biobank

| variant | Exposure allele | CARDIoGRAM ES | CARDIoGRAM SE | UK Biobank ES | UK Biobank SE |
|---|---|---|---|---|---|
| rs36005199 | G | -0.0570 | 0.0124 | -0.0204 | 0.0108 |
| rs55674565 | T | -0.0414 | 0.0097 | -0.0184 | 0.0101 |
| rs2883233 | C | -0.0434 | 0.0168 | -0.0093 | 0.0109 |
| rs117335915 | T | 0.0305 | 0.0348 | -0.0221 | 0.0147 |
| rs113201466 | G | -0.0382 | 0.0226 | -0.0153 | 0.0133 |
| rs62075782 | A | 0.00996 | 0.0151 | -0.0146 | 0.0117 |
| rs145940140 | T | -0.0730 | 0.0687 | -0.0241 | 0.0268 |
| rs143382920 | A | -0.0778 | 0.0650 | -0.0169 | 0.0298 |
| rs117981684 | G | -0.0566 | 0.0336 | 0.0321 | 0.0387 |
|  |  |  |  |  |  |
| Inverse-variance Meta-analysis |  | -0.0356 | 0.0058 | -0.0161 | 0.0046 |
|  |  |  |  |  |  |
|  |  | ES | SE | OR (95%CI) |  |
| Combined Cardiogram & UKBB |  | -0.0233 | 0.0035 | 0.977 (0.970-0.984) p=5.0E-11 |  |
|  |  |  |  |  |  |
| Adjusted per 10 mg/dl lower LDL-C |  | -0.1923 | 0.0293 | 0.825 (0.779-0.874) p=5.1E-11 |  |

NB: rs34349578 did not pass QC in UK Biobank, so rs36005199 was used as the nearest proxy ($r^2$=0.91)
NB: adjusted for 10 mg/dl in unweighted analyses using the summary effect per unit change in ACLY score of 1.21 mg/dl

**Table 8:** Individual *ACLY* score variants *weighted* associations with CAD per 10 mg/dl lower LDL-C in CARDIoGRAMplusC4D and UK Biobank

| variant | Conditional LDL-C ES | CARDIoGRAM ES | CARDIoGRAM SE | UK Biobank ES | UK Biobank SE |
|---|---|---|---|---|---|
| rs36005199 | -0.94 | -0.6064 | 0.1319 | -0.2168 | 0.1151 |
| rs55674565 | -1.25 | -0.3312 | 0.0776 | -0.1469 | 0.0808 |
| rs2883233 | -0.68 | -0.6382 | 0.2471 | -0.0888 | 0.1037 |
| rs117335915 | -1.83 | 0.1667 | 0.1902 | -0.1208 | 0.0801 |
| rs113201466 | -0.99 | -0.3859 | 0.2283 | -0.1547 | 0.1346 |
| rs62075782 | -0.98 | 0.1016 | 0.1541 | -0.1658 | 0.1324 |
| rs145940140 | -1.97 | -0.3706 | 0.3487 | -0.1224 | 0.1362 |
| rs143382920 | -1.19 | -0.6538 | 0.5462 | -0.1421 | 0.2500 |
| rs117981684 | -0.84 | -0.6738 | 0.4000 | 0.3822 | 0.4610 |
|  |  |  |  |  |  |
| Inverse-variance Meta-analysis |  | -0.3121 | 0.0537 | -0.1370 | 0.0388 |
|  |  |  |  |  |  |
|  |  | ES | SE | OR(95%CI) | |
| Combined Cardiogram & UKBB (per 10 mg/dl lower LDL-C) |  | -0.1997 | 0.0311 | 0.819 (0.770-0.871) p=1.5E-10 | |

**NB:** associations weighted by LDL-C effect size from conditional analyses as presented in Table 2, and scaled per 10 mg/dl lower LDL_C using ratio of ES method

When combined, the statistical evidence for an association between the *ACLY* score and the risk of CV events was robust (p = 1.5E-10 to 5.0E-11).

Furthermore, the *ACLY* score was associated with both major coronary events when analysed as a score calculated for each individual participant in the UK Biobank:

**Table 9:** Individual participant data unweighted and weighted *ACLY* scores association with MCVE in UK Biobank

| ACLY score | CAD ES | CAD SE | P value |
|---|---|---|---|
|  |  |  |  |
| Unweighted ACLY score above v below median | -0.0482 | 0.0139 | 9.33E-04 |
| (Adjusted per 10 mg/dl lower LDL-C for 2.51 mg/dl lower observed difference in LDL-C) | -0.1920 | 0.0552 | 9.33E-04 |
|  |  |  |  |
| OR (95% CI) per 10 mg/dl lower LDL-C: | 0.825 (0.741 -0.920) | | |
|  |  |  |  |
| Weighted ACLY score above v below median | -0.0469 | 0.0134 | 8.38E-04 |
| (Adjusted per 10 mg/dl lower LDL-C for 2.46 mg/dl lower observed difference in LDL-C) | -0.1905 | 0.0544 | 8.38E-04 |
|  |  |  |  |
| OR (95% CI) per 10 mg/dl lower LDL-C: | 0.827 (0.743 - 0.920) | | |

**NB:** associations weighted by LDL-C effect size observed for difference in LDL-C between participants with ACLY scores above and below median

The data above are those data that were available at the time and included in our manuscript. With the release of the updated outcomes data, the *ACLY* score remains robustly associated with the risk of both major cardiovascular events (MCVE) and major coronary events (MCE) in UK Biobank.

We note here that the definition and coding of MCE was performed by the UK Biobank Cardiometabolic working group (and not by the authors). The definition and coding of MCVE was simply MCE or ischemic stroke (a subset of the total stroke variable in UK Biobank). The variables ischemic stroke and MCVE were extracted and coded by the Cardiovascular Epidemiology Unit at the University of Cambridge and used across all studies using UK Biobank data for all investigators. Table 10 shows the associations of the ACLY genetic score with both MCVE and MCE in the most updated release of the UK Biobank data, as well as the associations of each variant included in the *ACLY* score with MCVE and MCE:

**Table 10:** Individual *ACLY* score variants associations with MCVE and MCE in UK Biobank

| variant | Exposure allele | Exposure allele frequency | MCVE ES | MCVE SE | MCVE P value | MCE OR | MCE SE | MCE P value |
|---|---|---|---|---|---|---|---|---|
| **rs36005199** | G | 0.779 | -0.0161 | 0.0098 | 9.9E-02 | -0.0192 | 0.0105 | 6.9E-02 |
| **rs55674565** | T | 0.379 | -0.0147 | 0.0090 | 1.1E-01 | -0.0192 | 0.0097 | 4.9E-02 |
| **rs2883233** | C | 0.130 | -0.0066 | 0.0108 | 5.4E-01 | -0.0079 | 0.0117 | 5.0E-01 |
| **rs117335915** | T | 0.978 | -0.0279 | 0.0234 | 2.3E-01 | -0.0293 | 0.0252 | 2.5E-01 |
| **rs113201466** | G | 0.950 | -0.0077 | 0.0205 | 7.0E-01 | -0.0170 | 0.0221 | 4.4E-01 |
| **rs62075782** | A | 0.223 | -0.0198 | 0.0104 | 5.7E-02 | -0.0141 | 0.0112 | 2.1E-01 |
| **rs145940140** | T | 0.991 | -0.0448 | 0.0523 | 3.9E-01 | -0.0464 | 0.0563 | 4.1E-01 |
| **rs143382920** | A | 0.993 | -0.0391 | 0.0334 | 2.4E-01 | -0.0241 | 0.0361 | 5.0E-01 |
| **rs117981684** | G | 0.979 | 0.0072 | 0.0326 | 8.3E-01 | 0.0360 | 0.0355 | 3.1E-01 |
| | | | | | | | | |
| **Inverse-variance Meta-analysis** | | | OR: 0.985 (0.976-0.994) | | P = 1.2E-03 | OR: 0.984 (0.975-0.994) | | P = 1.6E-03 |
| | | | | | | | | |
| OR per 10 mg/dl lower LDL-C weighted by marginal effects on LDL in UKBB | | | **OR: 0.710 (0.558-0.904)** | | P = 8.0E-03 | | | |
| | | | | | | | | |
| OR per 10 mg/dl lower LDL-C meta-analysis weighted by conditional effects on LDL in UKBB | | | **OR: 0.686 (0.519-0.907)** | | P = 5.4E-03 | | | |
| | | | | | | | | |

**NB:** rs34349578 did not pass QC in UK Biobank, so rs36005199 was used as the nearest proxy ($r^2$=0.91). Data from 351,267 participants of self-identified White ancestry not on current lipid lowering therapy, adjusted for age gender and the first five principal components

As compared to the analyses using the previous release of the data (and the analyses in our manuscript that were restricted to the 367,000 participants who self-identified as being of White British ancestry who were unrelated to all other participants as defined by a kinship coefficient > 0.044), the *ACLY* score remains associated with both MCVE and MCE in UK Biobank.

Table 11 below presents the associations of the *ACLY* score calculated for each participant in UK Biobank:

**Table 11:** Associations of *ACLY* genetic scores with MCVE and MCE in UK Biobank

| ACLY Score | MCVE OR (95% CI) | MCVE p value | MCE OR (95% CI) | MCE p value |
|---|---|---|---|---|
| Continuous variable scores | | | | |
| Unweighted score | 0.990 (0.982-0.997) | 6.7E-03 | 0.989 (0.981-0.997) | 8.5E-03 |
| Score weighted by conditional LDL ES (NEJM) | 0.990 (0.982-0.997) | 4.1E-03 | 0.989 (0.981-0.997) | 4.8E-03 |
| Score weighted by conditional LDL ES UKBB | 0.965 (0.940-0.991) | 1.0E-02 | 0.966 (0.939-0.993) | 1.9E-02 |
| Dichotomous scores as an instrument of randomization dividing participants with scores above and below median | | | | |
| Unweighted score | 0.958 (0.930-0.987) | 4.7E-03 | 0.957 (0.932-0.983) | 3.6E-03 |
| Score weighted by conditional LDL ES (NEJM) | 0.969 (0.946-0.993) | 1.1E-02 | 0.964 (0.939-0.989) | 5.4E-03 |
| Score weighted by conditional LDL ES UKBB | 0.972 (0.950-0.996) | 2.3E-02 | 0.972 (0.948-0.997) | 3.2E-02 |

**NB:** rs34349578 did not pass QC in UK Biobank, so rs36005199 was used as the nearest proxy ($r^2$=0.91). Data from 351,267 participants of self-identified White ancestry not on current lipid lowering therapy, adjusted for age gender and the first five principal components

Regardless of whether analysed as a continuous variable or a dichotomous instrument of randomization, and regardless of whether the scores were unweighted, weighted by the conditional weights used in our manuscript, or weighted by the conditional effect of each variant on LDL-C calculated from the UK Biobank lipid data, the *ACLY* score was associated with a lower risk of both MCVE and MCE (with effect sizes very similar to the effect sizes reported in our original manuscript).

We note here a methodologic consideration. Ultimately, Mendelian randomization is merely an instrument of randomization, without any further assumption. Therefore, the least biased analysis is to simply use the unweighted genetic score as an instrument of randomization to divide the sample into groups and measure the observed differences between LDL-C and cardiovascular events between the groups being compared. When these analyses are performed in UK Biobank, as shown in Tables 6, 9 and 11, the *ACLY* score is strongly associated with both lower LDL-C and lower CVD.

By contrast, adjusting the CVD effect size of each variant included in the score by the corresponding effect size of that variant on LDL-C (which is the conventional Wald ratio method) is potentially treacherous when analysing a score that by design is constructed to improve the association of variants **weakly associated with the exposure**. This is because in the Wald ratio method (and other similar methods of adjustment), the outcome effect size (and corresponding standard error) for each variant is divided by the exposure effect size for that variant (the so-called usual ratio of effect sizes method). From this explanation it becomes clear that the uncertainty captured the variance of the association between the variant and the outcome is incorporated into the "adjusted" effect estimate because both the effect size and the corresponding standard error are divided by the same number, thus preserving the measure of uncertainty. By contrast, the association between the variant and the exposure is *assumed* to be so robust that it does not have any practical variability. While this is true for variants strongly associated with exposures at genome-wide level of statistical significance where the SE and corresponding confidence intervals are exceedingly small, it is distinctly not the case when using variants weakly associated with the exposure, such as the weak associations between genetic variants in the *ACLY* gene region and LDL-C, where the confidence intervals for the associations with LDL-C are wide for most variants.

Therefore, "adjusting" the CVD effect size (and corresponding SE) for the variants included in the *ACLY* score by their imprecise estimate of effect size for LDL-C can potentially lead to unstable "adjusted" point estimates of effect that disagree when performed in different populations using different weights (where the weights can be quite different due to the relatively wide confidence interval of the association with the exposure).

Therefore, the most "fair" (unbiased) way to compare the effect sizes of the associations of the *ACLY* score in different populations, or combine the effect sizes across populations, is to use the unweighted genetic score as an instrument of randomization only (without further assumptions) and use the observed differences in LDL-C and CV events among the groups being compared to calculate the association of the *ACLY* score with LDL-C and risk of cardiovascular disease. These analyses are presented in Tables 6, 9 and 11 – and show that the "unweighted" *ACLY* score is robustly associated with LDL-C and CV events when used as an unbiased instrument of randomization (which has the least statistical power but also has the least potential bias).

Therefore, based on the analyses presented above, we conclude that the *ACLY* score that we constructed to be an instrument that mimics the effect of ACL inhibition is strongly associated with lower plasma LDL-C, concordantly lower plasma apoB (strongly suggesting that the score is instrumenting a mechanism that lowers LDL-C through the LDL receptor pathway), and a corresponding lower risk of cardiovascular events. Therefore, the overall three-stage approach we used to construct the score consisting of several variants weakly associated with the biomarker reflecting the pharmacologic action of ACL inhibition does appear to instrument the effect of ACL inhibition.

***Further evidence that the ACLY score is instrumenting the effect of ACL inhibition in UK Biobank***

As noted above, the *ACLY* score was designed to create an instrument that mimics the effect of ACL inhibition. Unfortunately, there is no current biological evidence that the variants included in our *ACLY* score are associated with changes in *ACLY* gene expression, or ACL structure or function. We note, however, that among the 250 variants that have been reported to be associated with *ACLY* expression in GTEx, only 3 are independently inherited and only one is weakly associated with plasma LDL-C (rs2304497 – a protein truncating variant within the ACLY gene that was reported to be weakly associated with lower LDL-C among 280K participants of exome analysis of the Global Lipids Genetics Consortium with a p value of 0.01).[22]

Release of the biomarker data from UK Biobank after the publication of our manuscript provides an opportunity to further evaluate whether our ACLY genetic score is instrumenting the effect of ACL inhibition.

Because the objective of our study was to construct an instrument that mimics the effect of ACL inhibition, the best biological evidence to make inferences about whether our genetic score may be instrumenting ACL inhibition are biochemical and clinical changes observed in randomized trials evaluating the inhibition of ACL by bempedoic acid.

In randomized trials, inhibiting ACL with bempedoic acid lowers LDL-C, concordantly lowers apoB (by inhibiting cholesterol biosynthesis leading to the up-regulation of LDL receptors), and is associated with lower body weight, lower HgbA1C, and lower diabetes risk.[12-15]

We noted above that the ACLY score is associated with both lower LDL-C and concordantly lower apoB in UK Biobank thus suggesting that the score is instrumenting a mechanism that lowers LDL-C through the LDL receptor pathway.

In addition, as shown in Table 12 below, like ACL inhibition with bempedoic acid, our *ACLY* score is also associated with lower weight, lower HgbA1C, and lower risk of diabetes in the UK Biobank.

Table 12: Association of unweighted and weighted *ACLY* scores with biomarkers and diabetes in UK Biobank

| Biomarker or phenotype | Unweighted score ES continuous | Unweighted score SE continuous | Unweighted score p continuous | Unweighted score ES dichotomous | Unweighted score SE dichotomous | Unweighted score p dichotomous |
|---|---|---|---|---|---|---|
| **Weight (kg)** | -0.061 | 0.013 | 4.6E-06 | -0.143 | 0.043 | 9.0E-04 |
| HgbA1C (%) | -0.035 | 0.006 | 1.5E-08 | -0.097 | 0.020 | 1.1E-06 |
| Diabetes | 0.990 | 0.003 | 1.8E-04 | 0.974 | 0.008 | 1.9E-03 |
| | (0.984-0.996) | | | (0.959-0.989) | | |

**NB:** associations unweighted scores treated as a continuous variable and as dichotomous instrument of randomization comparing participants with scores above v. below median. A total of 444,560 participants who self-identified white ancestry were included in the analysis, adjusted for age gender and the first five principal components. NB: rs34349578 did not pass QC in UK Biobank, so rs36005199 was used as the nearest proxy ($r^2$=0.91).

These data independently confirm the results that we included in our original manuscript, showing that the *ACLY* score was associated with lower weight (and lower BMI), and lower risk of diabetes using data from the GIANT and DIAGRAM consortia, respectively.[27,28]

Therefore, our **ACLY genetic score appears to have the same clinical profile as inhibiting ACL** with bempedoic acid.

Furthermore, as shown in Table 13, this clinical profile is a very unusual and is distinctly different than the clinical profile of variants in the HMG-CoA reductase gene that encodes the target of statins, and very different than the clinical profile of treatment with a statin – both of which have been shown to lower LDL-C and lower apoB concordantly, but to also INCREASE weight, INCREASE HgbA1C, and INCEASE the risk of diabetes.

Table 13: Comparison of *ACLY* score with Bempedoic acid (ACL inhibitor), *HMGCR* score, statins, and generic LDL-C score

| Genetic score of Therapy | LDL-C | apoB | CAD | weight | HgbA1C | Diabetes |
|---|---|---|---|---|---|---|
| **ACLY score** | decrease | decrease | decrease | decrease | decrease | decrease |
| ACL inhibitor (Bempedoic acid) | decrease | decrease | ? | decrease | decrease | decrease |
| | | | | | | |
| **HMGCR score** | decrease | decrease | decrease | increase | increase | increase |
| statins | decrease | decrease | decrease | increase | increase | increase |
| 102 variant LDL genetic score | decrease | decrease | decrease | increase | increase | increase |

This difference is biologically plausible because ACL is located upstream to HMG-CoA reductase proximal to the bifurcation of the lipid pathway into the cholesterol biosynthesis and intrinsic fatty metabolism pathways.

In addition, as shown in Table 13 above, the clinical profile of our *ACLY* score and ACL inhibition with bempedoic acid is also distinctly different than the clinical profile of a generic LDL-C score composed of 100 variants associated with LDL-C at genome-wide level of significance.  Like HMG-CoA reductase variants and statins, this generic LDL-C instrumental variable genetic score is also associated with lower LDL-C, concordantly lower apoB (because reducing LDL particles through the LDL receptor is the predominant biological mechanism for changes in plasma LDL-C), but an INCREASE in weight, INCREASE in HgbA1C, and an INCEASE the risk of diabetes.

 The difference in the clinical profile between our *ACLY* score and the generic LDL-C instrumental variable genetic score strongly suggests that our ACLY score is not instrumenting a generic LDL-C lowering mechanism, or some other gene in the region that may lower LDL-C through the LDL receptor pathway, but instead is instrumenting a very specific mechanism of lowering LDL-C.  We note that the only gene in this region that is known to be involved in lipid metabolism, and the only gene in the region known to code for a protein that when inhibited leads to lower plasma LDL-C with a clinical profile that is distinctly different than the clinical profile of other mechanism of lowering LDL-C is *ACLY*.

*ACLY score also mimics effect of an ACLY protein truncating mutation*

The release of the biomarker data from UK Biobank also allows a closer evaluation of the only protein truncating mutation in *ACLY* that is associated with lower LDL-C.  As noted above, this variant has been reported to be weakly associated with lower LDL-C (0.3 mg/dl, p=0.01) in the exome analysis of 280K participants in the Global Lipids Genetics Consortium, and associated with a concordantly (but non-significantly) lower risk of CAD (p=0.65) in the Myocardial Infarction Genetics and CARDIoGRAM Exome meta-analysis of Exome-chip studies including 42,335 patients and 78,240 controls of European descent.[29,30]

Table 14 shows that this protein truncating *ACLY* variant is associated with lower LDL-C (p=0.012), concordantly lower apoB (p=0.001), and a corresponding lower risk of major cardiovascular events (p=0.026) among participants enrolled in the UK Biobank.

Table 14: Associations of ACLY protein truncating variant rs2304497 with biomarkers and outcomes in UKBB

| Biomarker | Effect size (mg/dl) | SE | P value |
|---|---|---|---|
| LDL-C | **-0.337** | **0.1342** | **0.012** |
| apoB | **-0.313** | **0.0954** | **0.001** |
| Outcomes | OR (95% CI) | | P value |
| MCVE | **0.964 (0.934 - 0.995)** | | **0.026** |
| Diabetes | **0.981 (0.962 - 0.999)** | | **0.048** |

NB: associations of rs2304497 as dichotomous variable (carrier v. non-carriers). Analysis includes 364,128 unrelated UK Biobank participants self-identified as of white European ancestry, including 23,914 participants who experienced a first major cardiovascular event (defined as the first occurrence of fatal or non-fatal MI, ischemic stroke, coronary revascularization of CHD death).

Remarkably, however, Table 14 also shows that this *ACLY* protein truncating variant is also associated with a *lower risk of diabetes* – just like our *ACLY* score and ACL inhibition with bempedoic acid – but very different from *HMGCR*, *NPC1L1*, *PCSK9*, and *LDLR* variants, the 100-variant generic LDL-C instrumental variable genetic score, and treatment with a statin.

Therefore, our *ACLY* genetic score appears to have the same clinical profile as inhibiting ACL with bempedoic acid AND the same clinical profile as the protein truncating variant in *ACLY*. Together, these finding provide independent "genetic target validation" for *ACLY* as a therapeutic target.

Of course, it is possible that all these findings, replicated in multiple independent data sets, demonstrating a remarkably similar clinical profile of our *ACLY* score with the distinct clinical profile of ACL inhibition by bempedoic acid and decreased *ACLY* expression due to a protein truncating mutation in *ACLY*, could have occurred repeatedly by chance alone. However, that would be highly unlikely.

Therefore, we believe that it is reasonable to conclude that our *ACLY* genetic score does appear to instrument the effects of ACL inhibition, precisely what it was designed to do.

*Further evidence ACLY genetic target validation in UK Biobank*

The release of the biomarker data in the very large UK biobank also provides an opportunity to identify other rare variants within the *ACLY* gene that may be more strongly associated with LDL-C than common variants within and immediately around the *ACLY* gene. Despite an exhaustive, and computationally intensive, search over several weeks we were not able to identify any common or rare variants, either alone or in combination, that were strongly associated with LDL-C or that performed better than our *ACLY* score.

However, because the overwhelming majority of the 80 million imputed variants in the UK Biobank are rare variants (defined as having a MAF < 0.001), we were able to identify 64 independently inherited rare variants physically located within the *ACLY* gene itself (and not correlated with any other rare variant in the gene in individual participant analysis). The position of these variants within the *ACLY* gene are shown in the Figure:

Figure 1: **63 rarer variants (MAF < 0.005) within ACLY gene**

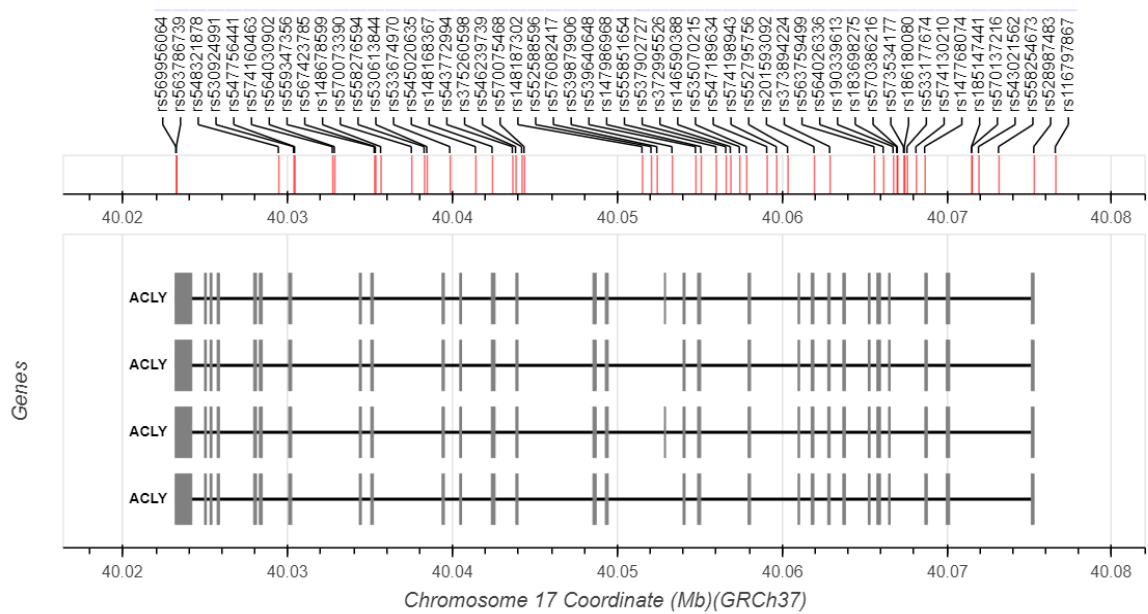

And their variant ID is provided in Table 15:

Table 15: **Score composed of 63 rarer variants (MAF < 0.005) within ACLY gene**

| variant | position | maf |
|---|---|---|
| rs569956064 | chr17:40023204 | 0.0008991 |
| rs563786739 | chr17:40023228 | 0.00067755 |
| rs548321878 | chr17:40029424 | 0.0004447 |
| rs530924991 | chr17:40030342 | 0.0008898 |
| rs547756441 | chr17:40030405 | 0.0000606 |
| rs574160463 | chr17:40032696 | 0.00025955 |
| rs574160463 | chr17:40032696 | 0.00025955 |
| rs564030902 | chr17:40032821 | 0.0003101 |
| rs559347356 | chr17:40035238 | 0.00038395 |
| rs567423785 | chr17:40035308 | 0.00002665 |
| rs148678599 | chr17:40035629 | 0.00272075 |
| rs570073390 | chr17:40037494 | 0.0021912 |
| rs558276594 | chr17:40038267 | 0.000186 |
| rs530613844 | chr17:40038429 | 0.00057075 |
| rs533674970 | chr17:40039824 | 0.0004665 |
| rs537210980 | chr17:40040601 | 0.000761 |
| rs545020635 | chr17:40041378 | 0.00408445 |
| rs148168367 | chr17:40042392 | 0.0003839 |
| rs543772994 | chr17:40043623 | 0.0005341 |
| rs375260598 | chr17:40043818 | 0.00031125 |
| rs546239739 | chr17:40044188 | 0.0000698 |
| rs570075468 | chr17:40044334 | 0.0001076 |
| rs148187302 | chr17:40051492 | 0.00005265 |
| rs552588596 | chr17:40052037 | 0.00017415 |
| rs576082417 | chr17:40052389 | 0.00093145 |
| rs539879906 | chr17:40053308 | 0.0005174 |
| rs539640648 | chr17:40054734 | 0.0002288 |
| rs147986968 | chr17:40055060 | 0.0001331 |
| rs534309713 | chr17:40055847 | 0.00009205 |
| rs555851654 | chr17:40055972 | 0.0001134 |
| rs537902727 | chr17:40056573 | 0.00058645 |
| rs372995526 | chr17:40056852 | 0.000046 |
| rs146590388 | chr17:40057400 | 0.0000166 |
| rs535070215 | chr17:40057818 | 0.0000561 |
| rs547189634 | chr17:40059061 | 0.0001693 |
| rs574198943 | chr17:40059639 | 0.00005395 |
| rs552795756 | chr17:40060324 | 0.00170425 |
| 17:40060658_GGAAA_G | chr17:40060658 | 0.0007278 |
| rs201593092 | chr17:40061926 | 0.00106985 |
| rs373894224 | chr17:40062879 | 0.00132455 |
| rs782020021 | chr17:40064533 | 0.0006678 |
| rs567286666 | chr17:40064835 | 0.001309 |
| rs563759499 | chr17:40065570 | 0.00029765 |
| rs564026336 | chr17:40066132 | 0.00117195 |
| rs190339613 | chr17:40066728 | 0.000602 |
| rs183698275 | chr17:40066946 | 0.00003925 |
| rs570386216 | chr17:40066986 | 0.00002215 |
| rs573534177 | chr17:40067366 | 0.00144025 |
| rs186180080 | chr17:40067409 | 0.00029155 |
| rs533177674 | chr17:40067573 | 0.00040585 |
| rs574130210 | chr17:40068116 | 0.000096 |
| rs574130210 | chr17:40068116 | 0.000096 |
| rs147768074 | chr17:40068646 | 0.00044255 |
| rs185147441 | chr17:40071485 | 0.00103795 |
| rs570137216 | chr17:40071496 | 0.00003725 |
| rs543021562 | chr17:40071913 | 0.00024415 |
| rs558254673 | chr17:40073128 | 0.00373955 |
| rs387907380 | chr17:40075176 | 0.0003477 |
| rs528987483 | chr17:40075268 | 0.0002518 |
| rs116797867 | chr17:40076575 | 0.0000564 |
| rs562369323 | chr17:40078478 | 0.0011736 |
| rs538819381 | chr17:40079774 | 0.00009025 |
| rs747583969 | chr17:40082520 | 0.0006775 |

As shown in Table 16, UK Biobank participants who carried one or more of these rare variants had an average of 1.3 mg/dL lower LDL-C (p<3.2E-08), and a corresponding 2-3% lower risk of major coronary events (p= 0.015) as compared to non-carriers.

Table 16: ACLY score consisting of 63 rarer variants within ACLY gene association with LDL-C and CAD in UKBB

| UK Biobank population ancestry | LDL-C es | LDL-C se | LDL-C p | CAD es | CAD se | CAD p |
|---|---|---|---|---|---|---|
| 367,643 unrelated, British, white ancestry | -1.2729 | 0.2302 | 3.2E-08 | -0.0759 | 0.0288 | 0.0147 |
| 459,322 self-identified white ancestry | -1.2846 | 0.2100 | 9.4E-10 | -0.0551 | 0.0263 | 0.0477 |

Although not compelling, and only marginally useful for attempting to "provide a biological context for interpreting the results of prior ACL inhibitor randomized trials, inform the design of future trials, and estimate the clinical effect of lowering plasma LDL-C by inhibiting ACL" (which was the purpose of our study), together with the data for the common *ACLY* protein truncating variant, these data do provide further "genetic target validation" for ACL inhibition as a therapeutic target.

**Conclusion**

The genes that encode the targets of most therapies do not have rare variants with large-effect or common variants with moderate effects on the biomarker reflecting the pharmacologic action of the corresponding therapy. Therefore, to provide "genetic target validation", to provide naturally randomized genetic evidence to help guide the evaluation and development of these therapies, or to provide a biological context for interpreting the results of randomized trials evaluating these therapies is challenging. Novel methods are being developed to combine multiple variants in the gene encoding the target of a therapy that are weakly associated with the biomarker reflecting the pharmacologic action of the therapy into a genetic score that can be used to "instrument" the therapy. This is an active area of research that is rapidly evolving. We provided one example of an approach to solve this important problem. Developing methods to create robust genetic instruments for a much greater number of potential therapeutic targets is critical for improving the role of Mendelian randomization in the drug discovery and development process.[31]